# The role of handbooks in knowledge creation and diffusion: A case of science and technology studies

*Staša Milojević[a], Cassidy R. Sugimoto[a], Vincent Larivière[b], Mike Thelwall[c], Ying Ding[a]*

[a] School of Informatics and Computing, Indiana University Bloomington 47405-1901, United States

[b] École de bibliothéconomie et des sciences de l'information, Université de Montréal, Montréal, QC. H3C 3J7, Canada

[c] School of Technology, University of Wolverhampton, Wolverhampton WV1 1LY, United Kingdom

E-mail addresses: smilojev@indiana.edu (S. Milojević), sugimoto@indiana.edu (C.R. Sugimoto), vincent.lariviere@umontreal.ca (V. Larivière), m.thelwall@wlv.ac.uk (M. Thelwall), dingying@indiana.edu (Y. Ding)

## Abstract

Genre is considered to be an important element in scholarly communication and in the practice of scientific disciplines. However, scientometric studies have typically focused on a single genre, the journal article. The goal of this study is to understand the role that handbooks play in knowledge creation and diffusion and their relationship with the genre of journal articles, particularly in highly interdisciplinary and emergent social science and humanities disciplines. To shed light on these questions we focused on handbooks and journal articles published over the last four decades belonging to the research area of Science and Technology Studies (STS), broadly defined. To get a detailed picture we used the full-text of five handbooks (500,000 words) and a well-defined set of 11,700 STS articles. We confirmed the methodological split of STS into qualitative and quantitative (scientometric) approaches. Even when the two traditions explore similar topics (e.g., science and gender) they approach them from different starting points. The change in cognitive foci in both handbooks and articles partially reflects the changing trends in STS research, often driven by technology. Using text similarity measures we found that, in the case of STS, handbooks play no special role in either focusing the research efforts or marking their decline. In general, they do not represent the summaries of research directions that have emerged since the previous edition of the handbook.

## 1 Introduction

Knowledge production and diffusion are cornerstones of the development of science, yet little is known about some of their aspects, such as the evolution of topics. Studies to understand topic evolution broadly fall into two categories: ethnographic (interviews with scientists) and bibliographic (analysis of scholarly documents). Most of the extant knowledge obtained using the bibliographic method is based on the analysis of a single genre at a time, and the one genre that dominates is the journal article. This bias is to a large degree the result of easy accessibility of bibliographic data for journal articles in electronic form. However, any particular genre presents an incomplete picture of the diversity of the scholarly communication landscape, and the reliance on journal articles, especially in the social sciences and humanities, which often eschew journal articles in favor of books and book chapters (Larivière, Archambault, Gingras, & Vignola Gagné, 2006), is unjustified. For a more complete view of how topics emerge, mature, and interact, it is desirable to take into account other genres from the rich ecology of scholarly communication.

Studies have suggested that different genres play different roles both in scholarly communication and in the practice of scientific disciplines (Bazerman, 1988; Swales, 2004). For example, textbooks have been identified as one of the signs of discipline formation (Lattuca, 2002; Lenoir, 1997), and monographs are important in the social sciences and humanities (Nederhof, 2006), as are edited volumes (e.g., (Engels, Ossenblok, & Spruyt, 2012; Hargens, 1991). Handbooks, on the other hand, do not have an extensive history of being studied to inform our understanding of disciplinary development. This seems to be an omission, given the cumulative and core status of handbooks (Fagerberg, Fosaas, & Sapprasert, 2012; Landström, Harirchi, & Åström, 2012; Martin, Nightingale, & Yegros-Yegros, 2012). Vickery (2000), for example, defines handbooks as "systematic accounts of what was known, with extensive references" (p. 150). These accounts have changed over time: early 19$^{th}$ century and 20$^{th}$ century handbook were usually written by a single author or group of coauthors and served educational purposes, having summarized the known knowledge within a field. Contemporary handbooks are primarily edited works with a set of selected contributed pieces.

Handbooks are particularly interesting as a genre as their goal is to encapsulate the core of a coherent subject area in order to act as a reference point for its researchers (Landström et al., 2012). Although they have been used in a few studies (e.g., Fagerberg et al., 2012; Kratus, 1993; Landström et al., 2012; Leming & Nelson, 1995; Martin et al., 2012; Randes, Hagen, Gottlieb, & Salvador, 2010), they have been largely overlooked as a bibliometric unit of analysis because they are fairly absent from indexes and, therefore, require manual work to prepare for the analysis (e.g., Fagerberg et al., 2012; Landström et al., 2012; Martin et al., 2012). This may be changing, however, with both the Web of Science and Scopus now indexing citations in selected books (e.g., Gorraiz, Purnell, & Glänzel, 2013) and with citation analysis of books also being possible to some extent using Google Books (Kousha & Thelwall, in press).

The goal of this study is to understand the role the handbooks play, especially in highly interdisciplinary emerging areas within social science and humanities, and their relationship with journal articles. To shed light on this question we will focus on four decades of handbooks and journal articles belonging to the research area of Science and Technology Studies (STS). Our definition of STS is broad and encompasses two traditions—the quantitative and qualitative—of the study of science and technology. The qualitative approach has appropriated the label *STS*, while the quantitative approach is frequently referred to as *scientometrics*. In this study we will use STS to designate both traditions, but will precede them with adjectives *qualitative* or *quantitative* when needed. Both research approaches share a main goal: the understanding of science and technology.

Qualitative STS is an interdisciplinary field that emerged in the 1970s and is "rooted in a variety of disciplines, including history, philosophy, sociology of science and technology, anthropology, cultural studies, critical theory, feminist theory, gender studies, and postmodern history" (Van House, 2004, p. 3). It has been defined as "an interdisciplinary field that is creating an integrative understanding of the origins, dynamics, and consequences of science and technology" (Hackett, Amsterdamska, Lynch, & Wajcman, 2008a, p. [1]). Many of the formative events for qualitative STS occurred in the 1970s: the start of publication of journal *Social Studies of Science* (SSS) in 1971, the foundation of The Society for Social Studies of Science (4S) in 1975, and the publication of the first STS handbook in 1977.

Quantitative STS began in the 1950s and 1960s (Spiegel-Rösing & Price, 1977; Leydesdorff & van den Besselaar 1997; van den Besselaar 2001), and was recognized by the name *scientometrics* in the 1970s. The name of the field is a translation of the term *naukometriya*, proposed by a Russian pioneer of quantitative studies of science, V.V. Nalimov (Nalimov & Mulchenko, 1971). Scientometrics can be defined as the "quantitative study of science, communication in science, and science policy" (Hess, 1997, p. 75). It adds "a quantitative focus on texts and communication to the interdisciplinarity of science and technology studies" (Leydesdorff & Milojević, 2015, p. 4) . The field has often been classified under the umbrella of library and information science (LIS) (e.g., Åström 2002, van den Besselaar & Heimeriks 2006), likely due to the use of citation analysis (a core scientometric method) in early library studies (e.g., Gross & Gross, 1927). However, scientometrics was recently shown to be both cognitively (Milojević, Sugimoto, Yan, & Ding, 2011) and socially (Milojević & Leydesdorff, 2013) distinct from general LIS. Scientometrics has undergone accelerated growth since the landmark publication of edited volume *Toward a Metric of Science: The Advent of Science Indicators* (Elkana, Lederberg, Merton, Thackray, & Zuckerman, 1978), which in its role for the development of the field could be considered to be the quantitative counterpart of the first qualitative STS handbook. The growth of quantitative STS research has led to the creation of the specialized journal *Scientometrics* in the late 1970s, and more recently *Journal of Informetrics* (2007), even though a large number of quantitative STS studies continues to be published in mainstream information science journal *Journal of the American Society for Information Science and Technology* (Milojević, Sugimoto, Yan, & Ding, 2011).

The present study itself uses quantitative methods to study STS, as have several previous studies. For instance, White and Griffith (1982) studied the intellectual development of qualitative STS as revealed by patterns of author co-citation of the literature of the field. They analyzed how 71 authors–identified primarily by using the index of the first STS handbook (Spiegel-Rösing & Price, 1977)—were interrelated as evidenced by journal publications in the period 1972-1980. This analysis pointed to the existence of two main cognitive domains within qualitative STS: social studies of science and science policy studies. Leydesdorff and van den Besselaar (1997) analyzed journal-to-journal citations for four STS journals: *Scientometrics*, *SSS, Research Policy,* and *Science, Technology & Human Values* (*STHV*) to understand "the differentiation of communication structures in STS" (p. 168). Similarly, Van den Besselaar (2000) studied the communication between STS journals to understand the state of STS as a field of study. Based on the analysis of the same four journals (*SSS, STHV, Scientometrics,* and *Research Policy*) he identified three areas that were becoming increasingly differentiated over time: (a) qualitative STS, (b) quantitative STS (scientometrics), and (c) S&T policy studies. The study furthermore focused on the social aspects of STS (i.e., authors and institutions). It found that while the social and cognitive relations between scientometrics and policy studies were similar, the social relation among the researchers and institutions within qualitative STS and policy studies were much stronger than the cognitive relationships among the documents produced in these areas.

Recently, Martin, Nightingale, and Yegros-Yegros (2012) studied the development of STS, in terms of the prominent papers, authors, and institutions, by analyzing the knowledge base (cited works, authors and title keywords) from five "authoritative handbooks comprised of expert reviews of STS" (p. 1183), i.e., the same five handbooks used in this study. They used the genre of handbooks as a springboard to identify influential STS literature belonging to other genres (primarily books). The authors identified a number of phases in the development of the field, and confirmed the separation between quantitative and qualitative STS.

Our paper utilizes multiple genres, extensive data (including the full text of handbooks) and new methods of analysis in order to arrive at a comprehensive and dynamic picture of the domain of STS and to examine the degree to which topics diffuse across genres. As far as we know this is the first analysis of STS that is based on the full text of five authoritative handbooks published over the span of thirty years. The analyses in this paper will exploit the fact that it is the text itself that most directly reflects the cognitive content of a document. Therefore, we will use words appearing in the text or titles and abstracts as proxies for the topics and concepts in order to investigate the relations within and across the genres. Alternative approaches, such as citation and co-citation analysis, provide meaningful structures for a discipline, but they assign and interpret topics post hoc. While equating vocabulary with topics represents a simplification, we nevertheless believe that it is warranted to study the changes in topics in this way, since even when certain terms or phrases are superseded by different ones, the change is not a simple matter of synonyms. In particular, scientific vocabulary signifies temporal and geographical focus, and changes in vocabulary are not mere development of the language, but carry deeper significance.

Based on the analysis of STS handbooks and journal articles we address the following questions: how similar are the handbooks among themselves? What have been the major topics covered in handbooks over time? What is the relationship between topics covered in journals and those that appear in handbooks? Do topics covered in handbooks lead to increased coverage in the journal literature, or does intensive study of topics results in their inclusion in handbooks?

Based on previous findings regarding the development of the STS and its scholarly communication channels (e.g., Martin et al. 2012), we expect that the analysis will confirm the qualitative-quantitative divide. In addition, we expect to see the gradual divergence of the two approaches when it comes to the topics they cover, similar to the divergence observed in terms of researchers, institutions and knowledge bases.

## 2 Data and methods

*Handbooks*

We were guided by two principles in the choice of handbooks: the handbooks had to cover both science and technology and they had to contain original chapters. These criteria resulted in the selection of five handbooks. The first one was published in 1977 under the title *Science, Technology and Society, A Cross-Disciplinary Perspective* (Spiegel-Rösing & Price, 1977) and is considered the first in the series of qualitative STS handbooks (Martin et al., 2012). The other two handbooks in the series were published "under the auspices of the Society for Social Studies of Science" (Martin et al., 2012, p. 1184) under the title of *Handbook of Science and Technology Studies* in 1995 (Jasanoff, Markle, Petersen, & Pinch, 1995) and 2008 (Hackett, Amsterdamska, Lynch, & Wajcman, 2008b), respectively. The first quantitative STS handbook, *Handbook of Quantitative Studies of Science and Technology* (van Raan, 1988), was published in 1988, and the second, *Handbook of Quantitative Science and Technology Research* (Moed, Glänzel, & Schmoch, 2005), in 2005. The same set of handbooks has been independently identified as authoritative for STS by Martin, Nightingale, and Yegros-Yegros (2012). Handbook titles, their abbreviations and the number of chapters are provided in Table 1.

Handbooks were scanned and converted to digital text using optical character recognition software. The full text of these handbooks, published between 1977 and 2008 and containing 136 chapters, includes 1.4 million occurrences of 28,500 unique words.

Table 1. Handbooks used in the study, listed chronologically.

| Handbook title | Abbreviation | Number of chapters | Number of pages |
|---|---|---|---|
| Rösing, I., & Price, D. J. d. S. (Eds.). (1977). Science, technology, and society: A cross-disciplinary perspective. London: SAGE. | H1-Qual77 | 15 | 607 |
| Van Raan, A. F. J. (Ed.). (1988). Handbook of quantitative studies of science and technology. Amsterdam: North-Holland. | H2-Quant88 | 21 | 774 |
| Jasanoff, S., Markle, G. E., Petersen, J. C., & Pinch, T. (Eds.). (1995). Handbook of science and technology studies. Thousand Oaks: SAGE. | H3-Qual95 | 28 | 820 |
| Moed, H. F., Glänzel, W., & Schmoch, U. (Eds.). (2005). Handbook of Quantitative Science and Technology Research. New York: Kluwer Academic Publishers. | H4-Quant05 | 34 | 800 |
| Hackett, E. J., Amsterdamska, O., Lynch, M., & Wajcman, J. (Eds.). (2008). The handbook of science and technology studies. Cambridge, MA: The MIT Press. | H5-Qual08 | 38 | 1065 |

*Journal articles*

One of the major challenges in using journal articles is that of disciplinary delineation. Our objective was to obtain the majority of articles that are relevant to STS without including too many that are not directly relevant. To this end we adopted a two-tiered procedure. First, we selected complete bibliographic records from Thompson Reuters Web of Science for *all* articles from journals that satisfy either of the criteria: (1) NSF subject category *Science studies* (56 journals), (2) journals cited at least five times in handbooks and not included in (1) (63 journals). This resulted in 109,164 articles published between 1956 and 2012. However, many of the articles selected in this way were clearly outside of the scope of STS (e.g., economics, general sociology, etc.). Therefore, we applied additional filtering in the following way. First, we identified *core journals* for both qualitative and the quantitative STS. For the former these are *SSS* and *STHV* (Leydesdorff & van den Besselaar, 1997; Martin et al., 2012). For the latter these are *Scientometrics* and *Journal of Informetrics* (Milojević & Leydesdorff, 2013). In addition, based on the findings from (Leydesdorff & van den Besselaar, 1997; Martin et al., 2012), we include *Research Policy* as another core journal, but its articles can be either quantitative or qualitative in nature. We used the criterion described below to classify the articles into those two categories. First, all articles from five core journals are kept for the final sample. For articles in other journals we retained only those that cited any of the five core journals[1]. They were classified as qualitative if the number of citations to two qualitative journals equaled or exceeded the number of citations made to two quantitative journals, and as quantitative in all other cases. The same classification was applied to articles from *Research Policy*, except that we kept (and classified as quantitative) even the articles that do not cite any of the four core journals. This filtering left 11,675 articles in the final sample, of which 4,104 were classified as qualitative and 7,571 were classified as quantitative. This two-tiered, citations-based approach has advantages over other methods. It has advantages over using keywords, especially in a field such as STS that uses terms that can appear as "general" words in a wide range of articles, in that it allows for inclusion of large number of articles with fairly high precision. In addition, the two-tiered approach has the advantages over using citation of core journals as the only criterion, since it takes into account that individual articles within a single journal may have different subjects than the journal itself.

Figure 1 shows the distribution of journal articles in each category by their year of publication. Publication volume in both groups has been rising throughout the period of coverage (since the late 1960s) and the rise has become especially rapid since the mid- 2000s. As will be discussed later, handbooks also split into qualitative and quantitative. The split gives the opportunity to test the above method for classification on handbook chapters and was found to be very reliable, coinciding with chapter placement (in qualitative vs. quantitative handbook) in all 72 cases (tested on the last two handbooks of each type).

Abstracts of journal articles are an important source of data for this study. However, the abstracts are not available for the entire time period for which we have bibliographic records, nor are they available for all articles. In our dataset the abstracts start appearing in 1991 and are available for 60% of the qualitative and 40% of the quantitative articles.

There were 93 different journals that published qualitative articles and 68 that published quantitative articles. Sixty-two journals contributed in both categories. Journals that published the greatest number of qualitative STS articles were: *Social Studies of Science*, *Science Technology & Human Values*, *Public Understanding of Science*, *Research Policy* and *Minerva* and these accounted for half of all the articles in this category. The five largest contributors in quantitative STS were *Scientometrics*, *Research Policy*,

---

[1] To determine citations we first identify all variants of core journal name abbreviations used in WoS.

*Journal of the American Society for Information Science and Technology (JASIST)*[2], *R & D Management*, and *Journal of Informetrics*—together these accounted for 80% of the quantitative articles.

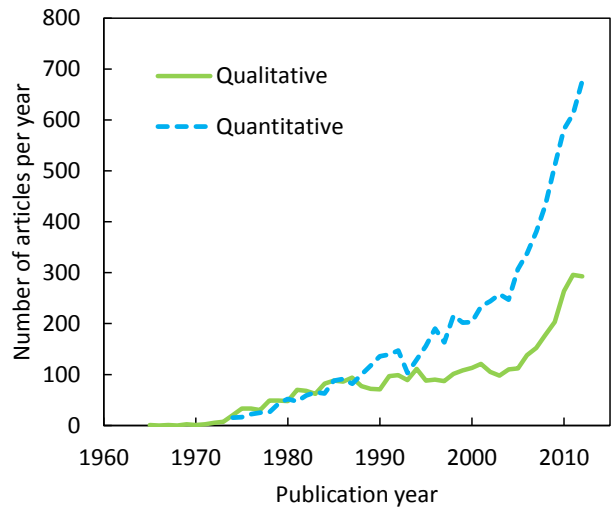

Figure 1. Number of journal articles in the area of science studies published between 1965 and 2012.

*Processing of text*

Once the data on handbooks and articles were collected we carried out the following procedure to identify terms (words and two-word phrases) that were used in the analysis. We first removed punctuation and non-word characters, and then consolidated word variants for plurals and excluded 707 general words. General words were identified by combining the lists of stop words supplied by WordStat (versions 5.1 and 6.1). We separately produced lists of frequencies of single words and of two-word phrases. The former was used in analysis of similarity, while the latter was used to identify frequent concepts.

*Measuring similarity based on text*

The method we used to quantify the similarity between two texts is the cosine similarity. Cosine similarity is an effective way of establishing the level of similarity among complex entities (Ahlgren, Jarneving, & Rousseau, 2003). This method measures the geometrical separation between multi-dimensional vectors, each representing some property, in this instance word occurrence frequency. The smaller the angle between the vectors (the closer the cosine is to 1) the more similar they are. If the two vectors are perpendicular (cosine = 0) then their attributes have nothing in common.

Based on the similarity matrix we then produced a diagram of hierarchical clusters (dendrogram) of items (handbooks or handbook chapters). Branches that split close to the root represent high-level clusters.

---

[2] The new name of the journal is *Journal of the Association for Information Science and Technology*

Branches that split farthest from the root represent entities that are most similar. Clustering is based on average-link clustering algorithm, which is a compromise between complete-link algorithm (which can be sensitive to outliers) and single-link algorithm (which tends to produce a long series of nested clusters). We have confirmed that the average-link algorithm more accurately clusters the items which are obviously similar based on titles alone than the other two methods.

To illustrate the performance of the similarity method we found, for three different handbook chapters, the most similar journal articles based on the words in abstracts and the titles. For example, for the handbook chapter: "Laboratory studies – the cultural approach to the study of science" by K. Knorr-Cetina the most similar articles based on the abstracts were: "In/visibilities of research: Seeing and knowing in STS" by L. Garforth, "Laboratizing and de-laboratizing the world: Changing sociological concepts for places of knowledge production" by M. Guggenheim, and "'Lab hands' and the 'Scarlet O': Epistemic politics and (scientific) labor" by P. Doing. The most similar articles based on titles were: "An R&D laboratory – case-study" by E.A. Wolff, "Science studies – what is to be done" by S. Restivo, and "Safe science: Material and social order in laboratory work" by B. Sims. We concluded that the word similarity method was able to identify items that intuitively appear to be similar.

# 3 Results

*Similarity of handbooks – the great divide*

We started with the question: can we quantitatively assess the level of similarity among the five handbooks? Do they reflect different traditions in STS? To answer these questions we calculated cosine similarity between the word frequencies of the main texts of all chapters in a given handbook, comprising of 490,349 occurrences of 28,456 different words. The results (Figure 2) show that handbooks are divided into two main branches. One contains handbooks H1-Qual77, H3-Qual95 and H5-Qual08 and the other contains H2-Quant88 and H4-Quant05. Within each main branch the handbooks have very high similarities (cosine ~0.9), while the similarity between the main branches is much lower (~0.55). It is sufficient to refer to the titles of handbooks (Table 1) to conclude that both H2-Quant88 and H4-Quant05 describe themselves as quantitative. Similarly, we will refer to H1-Qual77, H3-Qual95 and H5-Qual08 as qualitative handbooks.

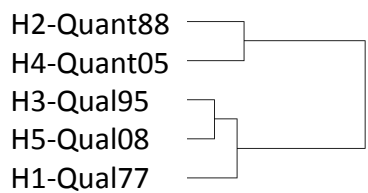

Figure 2. Dendrogram of the similarities between handbooks.

Our results correspond to the divide among handbooks found by Martin, Nightingale, and Yegros-Yegros (2012) as a result of cluster analysis based on a number of characteristics, such as thematic orientation of the handbooks, institutional affiliation of authors, impact, and keywords. In addition to the great divide, our results also suggest that the final qualitative handbook, H5-Qual08, represents a smaller departure from H3-Qual95 than the latter was with respect to H1-Qual77.

*Hierarchical clustering of handbook chapters*

We next explore the level of similarity between the 136 chapters contained in the handbooks. If we consider each chapter to represent a research topic within the STS, performing similarity analysis at the level of chapters allows us to map the structure of STS, where topics (chapters) that use similar vocabulary and therefore presumably address similar aspects of STS are grouped together in a hierarchical structure. The resulting dendrogram is shown in Figure 3.

The majority of chapters follow the qualitative – quantitative divide of the handbooks in which they were published, insofar that they all belong to one of the two high-level branches that split close to the root (the splitting point is noted as the diamond in Figure 3). This split is found regardless of the type of clustering algorithm used. There are only a few exceptions. First, there are two chapters, one from a qualitative handbook H5-Qual08 ("The right patients for the drug: Pharmaceutical circuits and the codification of illness") and one from a quantitative handbook H4-Quant05 ("Methodological issues of webometric studies") that are significantly different from all other chapters and they form their own high-level branches before the qualitative/quantitative divide. In addition, there are four chapters from quantitative handbooks that ended up in a predominantly qualitative branch (above the dashed line in Figure 3). These are: (a) two chapters from H2-Quant88 on technology ("The measurement of changes in technological output" and "Technological standards for research-intensive product groups and international competitiveness"), (b) one chapter from H4-Quant05 on science and technology in developing countries ("Science on the periphery: Bridging the information divide") and (c) one chapter from H4-Quant05 on gender ("Scientific and technological performance by gender"). There are no cases where a chapter from a qualitative handbook was placed in the quantitative branch. The chapters that were clustered among chapters of a differing approach signal topics that most strongly connect the two traditions.

There are several pairs of chapters that appear to be very similar to each other (i.e., they branch together far from the root). All such pairs are in the qualitative cluster. The most similar pair comes from H1-Qual77 and H5-Qual08: "Criticisms of science" and "The social study of science before Kuhn". Both provide an overview of pre-1960s studies of science. The next three pairs all seem to be updates of chapters that have already covered the topics of: (a) the relationship between science, technology and the military (H3-Qual95 and H5-Qual08: "Science, technology, and the military" and "Science, technology, and the military: Priorities, preoccupations, and possibilities"), (b) gender (H3-Qual95 and H5-Qual08: "Women and scientific careers" and "The coming gender revolution in science"), and (c) policy (H1-Qual77 and H3-Qual95: "Science policy studies and the development of science policy" and "Changing policy agendas in science and technology"). None of the authors of these chapter pairs are in common. Although not as similar as the above chapters, there have been quite a few instances of the updated chapters on the similar topics, e.g., laboratory studies, public understanding of science, science indicators, mapping of science, and technology indicators based on patents. From this we conclude that while subsequent editions of handbooks largely address aspects that were not covered previously, in some cases they merely provide updates.

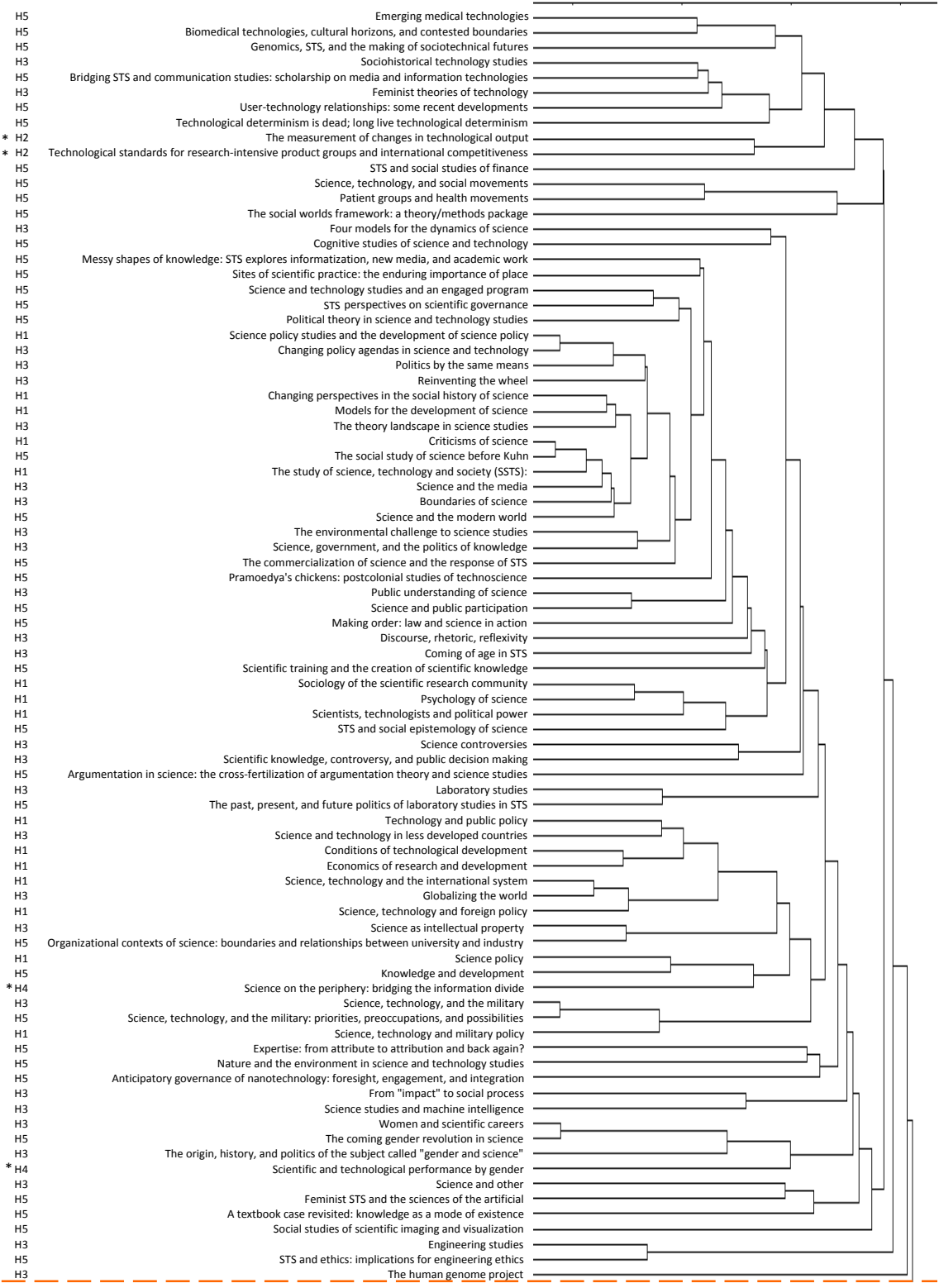

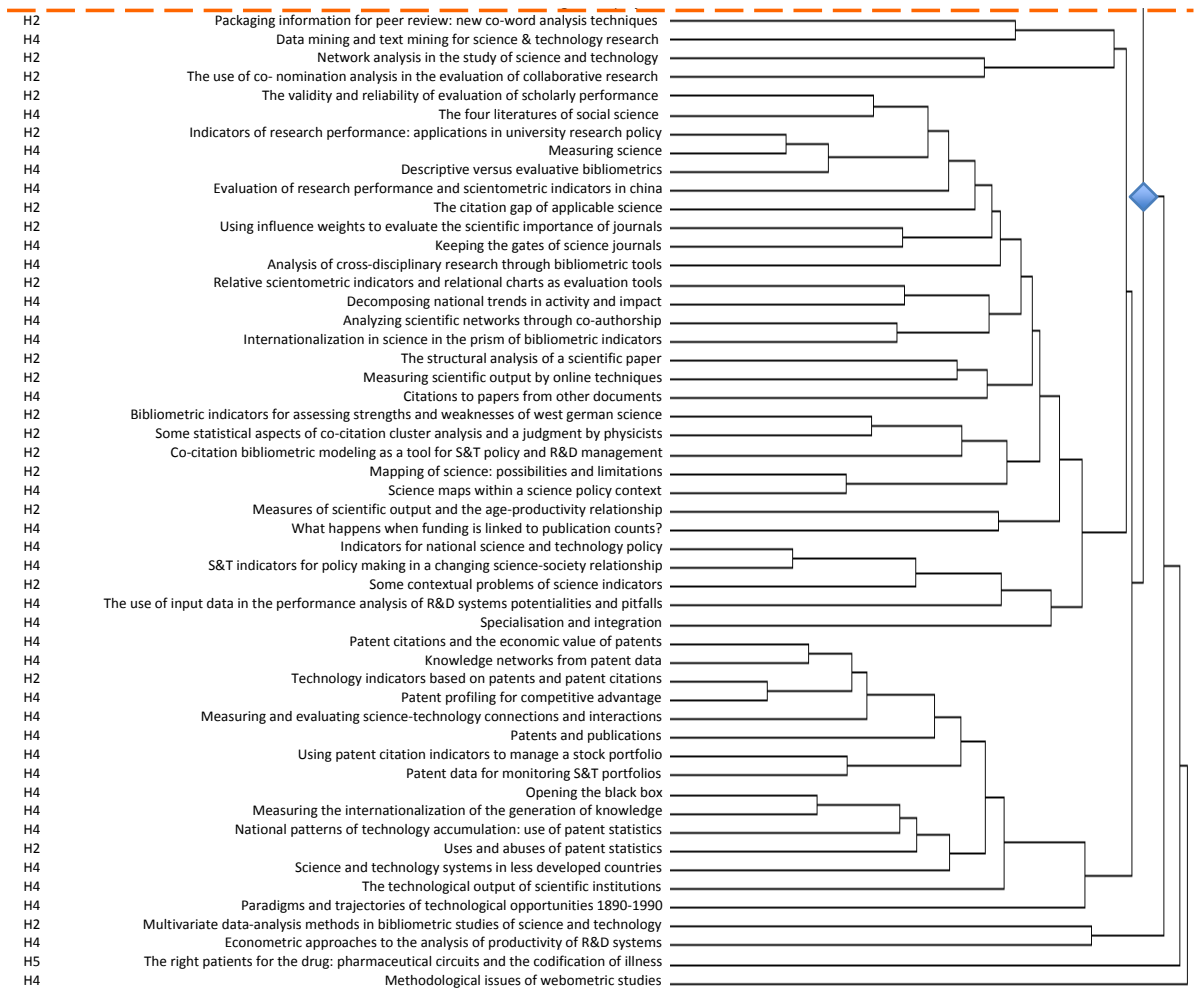

Figure 3. Hierarchical clustering of handbook chapters based on cosine similarity. The diamond represents the splitting point between branches that mostly contain chapters belonging to qualitative or quantitative handbooks (above and below the dashed line). Exceptions to this demarcation are few and are denoted with an asterisk at the leftmost edge. The bottom two chapters are outliers that separate above the main splitting point.

*Differences in approach between qualitative and quantitative STS*

Hierarchical clustering of handbook chapters revealed some topics that are of interest to both traditions of STS. These topics present excellent test beds to examine the commonalities and differences in how qualitative and quantitative STS approach such subjects. Two such subjects and the associated clusters of chapters were of particular interest: the one on technology that had five chapters from qualitative STS handbooks ("Sociohistorical technology studies", "Bridging STS and communication studies: scholarship on media and information technologies", "Feminist theories of technology", "User-technology relationships: some recent developments", "Technological determinism is dead; long live technological determinism") and two from quantitative STS handbooks ("The measurement of changes in technological output" and "Technological standards for research-intensive product groups and international competitiveness"). The other shared subject is the role of gender: there were three chapters from qualitative handbooks ("Women and scientific careers", "The coming gender revolution in science", "The origin, history, and politics of the subject called "gender and science"") and one from a quantitative handbook ("Scientific and technological performance by gender").

When it comes to studying technology, the quantitative handbooks focused on indicators and measurement (e.g., the fifth and sixth most common terms in these two handbooks were "indicator" and "measurement" respectively, and phrases such as "indicator values" and "technical indicators" were among the top phrases). These handbooks also focused on individual products (e.g., technologies such as "sensor systems", "laser diodes", "industrial robots"). On the other hand, chapters in qualitative handbooks were focused on technology as part of a larger system in which users and the social context play important role. Thus, the most commonly used words are "social" and "users" and phrases "technological determinism", "social shaping", "sociotechnical ensembles", "actor network", and "social construction". The qualitative handbooks also do not focus on individual products, but on a type of technology, such as "information technology". These findings reflect increased concern by qualitative STS scholars "about the impact of S&T developments on health, safety, and fundamental human values" (Jasanoff, 2010, p. 195).

Let us now look at the second common topic, gender. The focus of the qualitative handbooks is on "women scientists", their careers and gender differences. Quantitative handbooks focus primarily on "women authors" and "women inventors", that is, on particular activities that women engage in as scientists. The qualitative STS handbooks frequently refer to "feminist theory", while the quantitative chapter does not explicitly mention any theory. The choice of words to describe gender differences is also differentiating. One of the top phrases in the quantitative handbook chapter is "gender equality". While the qualitative handbooks use terms "equality" and "inequality" almost equally (12 versus 11 times), the quantitative handbook chapter has a significant bias towards "equality", using "equality" six times and "inequality" only once. The distinction between the uses of gendered terms is also revealing because the authors of chapters in qualitative handbooks position their discussion around both genders. They use term "women" 495 times and "men" and "male" 138 and 23. The quantitative study obviously does not juxtapose women to men to that extent. The handbook uses term "women" 62 times and "male" and "men" 10 and 4 times respectively. No chapter uses the term "female".

*Top topics – differences and similarities between the genres*

To better understand these similarities and differences we examined the fifteen most frequently occurring phrases and their trends from one handbook edition (of a given type) to another. We then extended the topic analysis to journal articles as well. We analyzed phrases rather than individual words since they provided a better context for some of the commonly occurring terms.

Table 2. Most frequent phrases in handbooks. Frequent phrases that appear in more than one handbook are shown in bold.

| H1-Qual77 | H2-Quant88 | H3-Qual95 | H4-Quant05 | H5-Qual08 |
|---|---|---|---|---|
| **SCIENCE POLICY** | CO CITATION | **SCIENTIFIC KNOWLEDGE** | PATENT APPLICATIONS | **SCIENTIFIC KNOWLEDGE** |
| **DEVELOPING COUNTRIES** | BIBLIOMETRIC MODELS | **SCIENCE STUDIES** | WEB PAGES | SOCIAL WORLDS |
| SCIENTIFIC DEVELOPMENT | DATA BASES | **SCIENCE POLICY** | **DEVELOPING COUNTRIES** | PATIENT GROUPS |
| FOREIGN POLICY | **HIGHLY CITED** | **INTELLECTUAL PROPERTY** | FRASCATI MANUAL | SOCIAL MOVEMENTS |
| **SCIENTIFIC COMMUNITY** | CO WORD | LABORATORY STUDIES | CO AUTHORSHIP | TECHNOLOGICAL DETERMINISM |
| **WAR II** | CITATION CLUSTER | **SCIENTIFIC COMMUNITY** | **HIGHLY CITED** | HEALTH MOVEMENTS |
| **SCIENTIFIC KNOWLEDGE** | **SCIENCE POLICY** | **SOCIAL STUDIES** | SOCIAL SCIENCE | INFORMATION TECHNOLOGIES |
| TECHNOLOGICAL DEVELOPMENT | FOREIGN TRADE | TECHNOLOGY STUDIES | PATENT CITATIONS | TWENTIETH CENTURY |
| NINETEENTH CENTURY | PEER REVIEW | **WAR II** | TECHNOLOGICAL OPPORTUNITIES | STS SCHOLARS |
| **SOCIAL STUDIES** | PATENT STATISTICS | TECHNOLOGY POLICY | INTERNATIONAL COLLABORATION | **SCIENCE STUDIES** |
| ECONOMIC GROWTH | SUBJECT AREAS | KNOWLEDGE SYSTEMS | PATENT DATA | **INTELLECTUAL PROPERTY** |
| **SCIENCE TECHNOLOGY** | PATENT SYSTEM | MILITARY TECHNOLOGY | **SCIENCE POLICY** | E SCIENCE |
| INTERNATIONAL AFFAIRS | WEST GERMANY | SCIENCE COMMUNICATION | JOURNAL ARTICLES | COLD WAR |
| ATOMIC ENERGY | CITATION INDEX | SOCIAL CONSTRUCTION | CITATION IMPACT | KNOWLEDGE PRODUCTION |
| SOCIAL HISTORY | GENETIC ENGINEERING | TECHNOLOGY ASSESSMENT | **SCIENCE TECHNOLOGY** | **SCIENCE TECHNOLOGY** |

The most frequently occurring phrases in the handbooks are given in Table 2. We see that despite the high formal degree of similarity between H2-Quant88 and H4-Quant05 on the one hand and H1-Qual77, H3-Qual95 and H5-Qual08 on the other, each handbook has a specific focus. It is interesting that most of the top 15 phrases in the latest qualitative handbook (H5-Qual08) did not appear in top 15 in the first (H1-Qual77). As for the two scientometrics handbooks, it is notable that the phrase 'bibliometric models' has disappeared from top phrases in H4-Quant05, as well as 'data base' and 'citation cluster'. On the other hand, H4-Quant05 debuts phrases such as 'web pages', 'open access', and 'knowledge flows'. We will discuss trends in the topic coverage in a separate section below.

**Table 3. Top 15 topics in qualitative STS handbooks and journals. The terms that occur frequently both in handbooks and article abstracts are bold.**

| Handbooks (full text) | Articles (abstracts) |
|---|---|
| **SCIENCE POLICY** | **SCIENTIFIC KNOWLEDGE** |
| **SCIENTIFIC KNOWLEDGE** | **TWENTIETH CENTURY** |
| **SCIENCE STUDIES** | TECHNOLOGY STUDIES |
| DEVELOPING COUNTRY | **NINETEENTH CENTURY** |
| SOCIAL MOVEMENT | **SCIENCE STUDIES** |
| **SCIENTIFIC COMMUNITY** | **SCIENCE POLICY** |
| **SOCIAL SCIENCE** | **SCIENTIFIC COMMUNITY** |
| TECHNOLOGICAL DEVELOPMENT | PUBLIC ENGAGEMENT |
| INTELLECTUAL PROPERTY | **SOCIAL SCIENCE** |
| SOCIAL WORLDS | KNOWLEDGE PRODUCTION |
| **TWENTIETH CENTURY** | COLD WAR |
| WORLD WAR II | SCIENTIFIC PRACTICE |
| SCIENCE & TECHNOLOGY | ACTOR NETWORK |
| PATIENT GROUP | PUBLIC PARTICIPATION |
| **NINETEENTH CENTURY** | STEM CELL |

We focus next on the top topics in all handbooks of a given tradition (from full text) as opposed to those in articles of the same tradition (based on abstracts). Top topics in qualitative STS exhibit a higher overlap between handbooks and articles than those in quantitative STS. For example, seven of the top fifteen terms appear in both handbooks and articles (Table 3). Two are related to particular time periods in the development of science (19$^{th}$ century and 20$^{th}$ century), and five are about particular topics (scientific knowledge, science studies, science policy, scientific community, and social science). The qualitative STS handbooks and articles have somewhat different foci. The handbooks seem to focus more on World War II (or at least make more references to it), and articles on the period following it, namely the Cold War. The handbooks include references to 'technological development' and articles focus on 'technology studies'. The handbooks employ 'social worlds' theory, while articles tend to use 'actor network' theory. The handbooks are interested in 'social movement' and 'patient group', while articles talk about 'public engagement' and 'public participation'. The articles are also focused on two of the major themes in qualitative STS 'knowledge production' and 'scientific practice'. Articles tend to write more about recent lines of research, e.g., 'stem cells'. Regardless of these small differences, there is high agreement in the topical foci of these genres. We'll discuss these further in the section on trending topics.

The top topics in quantitative STS exhibit a somewhat lower level of overlap between handbooks and articles (Table 4). In particular, only five out of fifteen top topics are shared. Out of those five phrases one is on the major tool used in scientometric analysis, 'citation index', two are on the types of analysis: 'co-citation' and 'co-authorship'. The handbooks have exhibited a higher interest in different aspects of patents, with three out of the top fifteen topics being on patents. Articles have stronger focus on research impact (e.g., impact factor, journal impact, citation impact). The *h*-index has also been one of the most studied indicators (Rousseau, Garcia-Zorita, & Sanz-Casado, 2013). However, it was introduced a year after the publication of the last quantitative STS handbook (Hirsch, 2005), explaining its absence from the handbooks. It is interesting that qualitative and quantitative handbooks share an interest in "developing countries", a topic which is not seen frequently in either qualitative or quantitative journals. A possible explanation for this is that journal articles focus on individual countries whereas handbook chapters give a broad overview of many studies of individual developing countries. Alternatively, the focus on the developing countries in handbooks may be the result of different nature of the genres, such that handbook editors could have invited the authors to specifically cover this topic deemed important by them.

Table 4. Top 15 topics in quantitative STS handbooks and journals. The terms that occur frequently both in handbooks and article abstracts are bold.

| Handbooks (full text) | Articles (abstracts) |
| --- | --- |
| **CO CITATION** | IMPACT FACTOR |
| BIBLIOMETRIC MODEL | H INDEX |
| **SOCIAL SCIENCE** | **CITATION INDEX** |
| CITATION COUNT | **SOCIAL SCIENCE** |
| **HIGHLY CITED** | **CO AUTHORSHIP** |
| DATA BASE | INTERNATIONAL COLLABORATION |
| DEVELOPING COUNTRY | **CO CITATION** |
| PATENT CITATION | BIBLIOMETRIC INDICATOR |
| PEER REVIEW | **HIGHLY CITED** |
| SCIENCE POLICY | CITATION RATE |
| PATENT APPLICATION | CITATION IMPACT |
| PATENT OFFICE | JOURNAL IMPACT |
| **CITATION INDEX** | SCIENTIFIC JOURNAL |
| SUBJECT AREA | SCIENTIFIC OUTPUT |
| **CO AUTHORSHIP** | SCIENTIFIC PRODUCTION |

*Trending topics*

The previous analyses provide an overview of the intellectual territory covered by two major STS approaches over the last four decades. However, these analyses provide only snapshots of the nature of the field. They do not elucidate knowledge creation dynamics or provide an understanding of the diffusion of topics. Such trends are naturally present in journal articles which reflect the interests and the state of knowledge at a given time, but they are also present in handbooks, because they also, to some extent, present a summary of the state of the field at a given time (as most explicitly evidenced by many instances of chapters that update the same topics in previous handbooks). For example, some terms (such as 'data base' in H2-Quant88) have disappeared from usage in the later editions of STS handbooks, and yet feature prominently on the list of top terms due to its cumulative nature. To get a more nuanced view of the field we need to analyze the trends in the usage of words and phrases over time. We focus on two trends: rising and declining in usage of terms (in both genres). We define rising and declining topics as frequent phrases whose usage has changed by more than 50% (which includes phrases that did not exist at one time but came into usage at a later time period). As above, we analyze qualitative and quantitative STS separately.

We examine trending topics for both handbooks and contemporaneous journal articles. For the analysis of trending topics in qualitative STS handbooks, we determine the difference in topics usage between 1995 and 2008 by analyzing the frequency of phrases as they occur in the full text of handbook chapters published in these two years (handbook H3-Qual95 and H5-Qual08). For the analysis of trending topics in qualitative STS journal literature we examine the change in topic usage between two five-year periods immediately preceding the publication of respective handbooks, 1991-95 and 2004-08, by analyzing the frequency of phrases as they occur in the abstracts of articles published in these time periods.

Table 5. The top declining topics in qualitative STS. The terms that are in common both in handbooks and article abstracts are bold.

| Handbooks (full text) between 1995 (H3-Qual95) and 2008 (H5-Qual08) | Articles (abstracts) between 1991-95 and 2004-08 |
|---|---|
| SCIENTIFIC KNOWLEDGE | **SCIENCE STUDIES** |
| **SCIENCE STUDIES** | **SCIENCE POLICY** |
| **SCIENCE POLICY** | **SCIENTIFIC COMMUNITY** |
| LABORATORY STUDIES | **SOCIAL STUDIES** |
| KNOWLEDGE SYSTEM | SOCIAL CONSTRUCTION |
| **SCIENTIFIC COMMUNITY** | GENETIC ENGINEERING |
| TECHNOLOGY POLICY | ACTOR NETWORK |
| TECHNOLOGICAL DEVELOPMENT | WAR II |
| **SOCIAL STUDIES** | UNITED KINGDOM |

Let us start with the declining topics in qualitative STS (Table 5). Both the handbooks and the articles have seen a decline in four topics: science studies, science policy, scientific community and social studies. We see the decline in interest in "laboratory studies", "technology policy" and "technological development". We also see a decline in the "social construction" and "actor network" phrases. Both of these stand for two major approaches to theorizing science and technology that have been highly influential in the 1980s and 1990s (Sismondo, 2008; Van House, 2004). In addition, we see a decline in the interest in "genetic engineering". We have also witnessed a decrease of interest in the WWII period and the United Kingdom in the journal literature.

Next we examine the top rising topics in qualitative STS (Table 6). Although there is no overlap in the exact phrases among the rising topics for qualitative STS handbooks and the journal literature, we still see some commonalities in terms of the social approach to studying science, exemplified by Knorr Cetina's work, and a new focus on the production and role of science in the modern world. We also see more retrospective discussion of 'twentieth century' and 'cold war'.

**Table 6. The top rising topics in qualitative STS**

| Handbooks (full text) between 1995 (H3-Qual95) and 2008 (H5-Qual08) | Articles (abstracts) between 1991-95 and 2004-08 |
| --- | --- |
| SOCIAL MOVEMENT | KNOWLEDGE PRODUCTION |
| SOCIAL WORLDS | PUBLIC PARTICIPATION |
| HEALTH MOVEMENT | SOCIAL SCIENCE |
| INFORMATION TECHNOLOGY | COMPUTER SCIENCE |
| TECHNOLOGICAL DETERMINISM | RISK ASSESSMENT |
| TWENTIETH CENTURY | POLICY MAKERS |
| STS SCHOLAR | PEER REVIEW |
| KNORR CETINA | SCIENTIFIC INSTITUTIONS |
| E SCIENCE | MEDICAL PRACTICE |
| COLD WAR | INTELLECTUAL PROPERTY |

In order to examine the trending topics within quantitative STS, we analyzed the full text of quantitative handbook chapters between the 1988 and 2004 editions. We could not examine the corresponding trending topics in the journal literature, since article abstracts are not available prior to 1991. By examining the trending topics (Tables 7 and 8) we see that the usage of 'co-citation' and 'co-word' has declined, while the usage of 'co-authorship' and a related topic of 'international collaboration' have risen. The decline in interest in 'science policy' matches that of qualitative STS. It is interesting that there is a decline of interest in 'peer review', while the same topic has experienced an increase in interest within the qualitative STS journal literature. The increased interest in patents and the implications of technology advancement are also highly visible. In particular, there has been an increase of usage of phrases: 'patent office', 'patent citation', 'patent data' and 'technological opportunity' as well as 'knowledge flow'. The increase of interest in 'knowledge flow' might also be tied to the interest in the academia-industry transfer of knowledge.

**Table 7. The top declining topics in scientometrics**

| Handbooks (full text) between 1988 and 2004 |
|---|
| CO CITATION |
| DATA BASE |
| SUBJECT AREA |
| HIGHLY CITED |
| CO WORD |
| PEER REVIEW |
| INNOVATIVE ACTIVITY |
| SCIENCE POLICY |
| FOREIGN TRADE |
| PATENT STATISTICS |

**Table 8. The top rising topics in scientometrics**

| Handbooks (full text) between 1988 and 2004 |
|---|
| DEVELOPING COUNTRY |
| PATENT OFFICE |
| PATENT CITATION |
| KNOWLEDGE FLOW |
| CO AUTHORSHIP |
| FRASCATI MANUAL |
| KNOWLEDGE BASE |
| TECHNOLOGICAL OPPORTUNITY |
| INTERNATIONAL COLLABORATION |
| PATENT DATA |

*The knowledge base and its usage in different genres in STS*

Another way to compare qualitative and quantitative STS in different genres is to examine their usage of the journal literature. We focused only on the usage of four core journals, which account for the bulk of references, both in handbooks of a given tradition and in the respective body of journal articles (Figure 4). For more extensive analysis of references in the five handbooks by and their usage in literature through citations, see Martin, Nightingale, and Yegros-Yegros (2012).

Qualitative STS draws mostly from qualitative core journals and quantitative STS from scientometrics core journals. Both handbooks and journal articles in STS highly utilized articles from *SSS* and *STHV*, and de-emphasized quantitative journals such as *Research Policy* and *Scientometrics*. The inverse was true for the quantitative handbooks and journals. However, quantitative handbooks used *SSS* more than qualitative STS (either handbooks or articles) used articles from the journal *Scientometrics*. We did not examine the actual articles that were referenced. However, in the earlier period *SSS* published influential quantitative works. Thus, it is plausible that quantitative researchers actually cite scientometric literature published in *SSS*. It is notable that quantitative STS literature, especially journal articles, rarely use any of the research published in *STHV*.

In addition, we found that quantitative STS, both handbooks and articles, draws much more extensively from *Research Policy* than does qualitative STS. This is in agreement with Fagerberg, Fosaas, and Sapprasert's (2012) finding that while the researchers in the field of innovation studies, whose leading journal is *Research Policy,* extensively use STS literature, the reverse is not the case. Both qualitative STS genres use *SSS* more than any other core journal. Handbook chapters used *STHV* articles much more extensively than did qualitative journal literature. Both qualitative STS handbooks and articles use *Scientometrics* very sparingly. There is a somewhat larger usage of *Research Policy*. Quantitative STS handbooks and articles use *Research Policy* much more extensively, with journal literature using it even more than the handbooks.

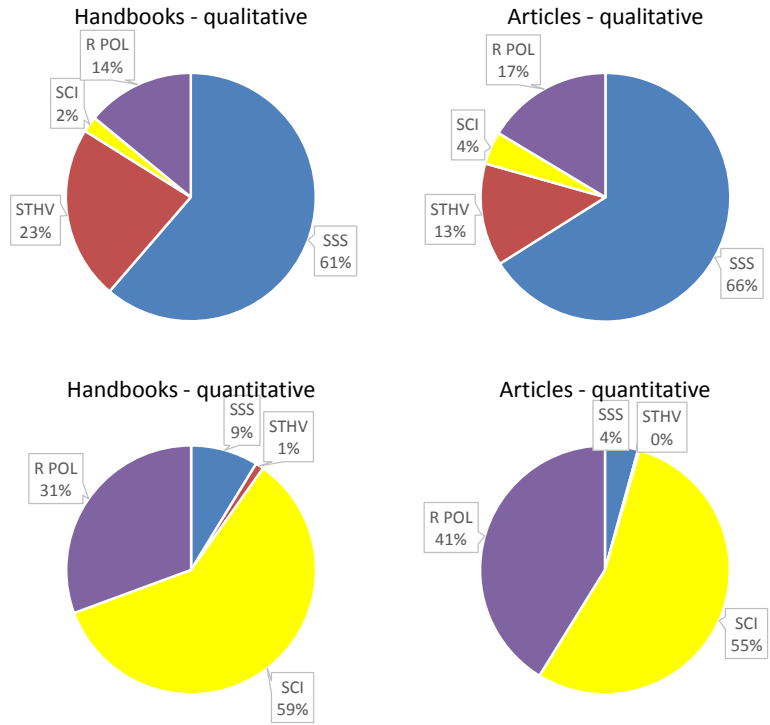

**Figure 4. Usage of specific core journals as a fraction of total core journal references in qualitative and quantitative STS of both genres. Journal abbreviations:** *Scientometrics* **(SCI),** *Social Studies of Science* **(SSS),** *Research Policy* **(R POL)*,* and *Science, Technology & Human Values* **(STHV).**

*The role of handbooks in knowledge production*

In this section we focus on the relation between handbooks and journal articles. In particular, we are interested in finding whether the topics that have been covered in the handbooks were more "popular" before or after the publication of handbooks. Specifically, we wish to determine whether handbook topics (i.e., different chapters) play a role in focusing research interest, or whether they provide summaries of mature or even declining topics. Answering these questions can help us understand whether the observed trends reflect changes in research activity within STS as a field, or just the topic coverage of the handbooks.

To this end we calculated cosine similarity between each of the 136 handbook chapters with titles of all articles in a given category (qualitative or quantitative) published over the 40 year period. We then averaged these similarities over chapters in a given handbook in each year. The averaging is performed only on the top 10% of articles that are most similar, in a given year, to a given handbook. This analysis allowed us to assess the degree to which the topic of handbook matched those of the journal articles over the period of four decades. Note that this measure is relative, so it is insensitive to the overall rise in STS literature (Figure 1). The results are shown in Figure 5.

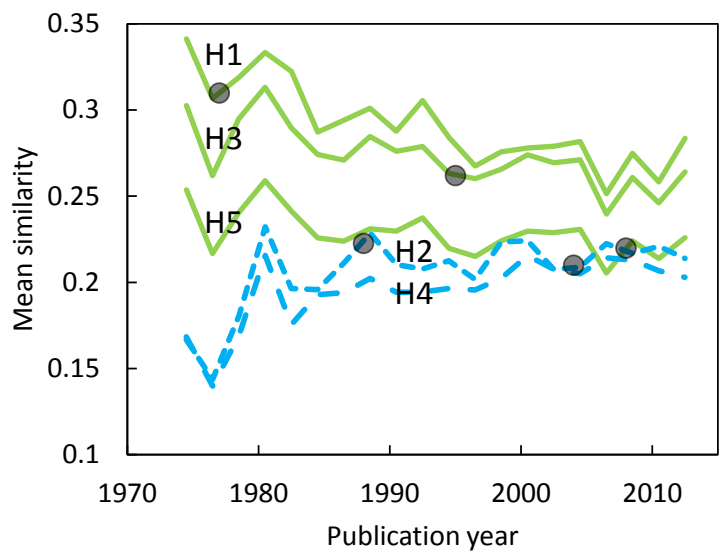

Figure 5. Trend of mean similarity of 10% of articles that are most similar in each year to chapters in a given handbook. Qualitative literature is shown by solid green lines and quantitative by dashed blue lines.

The graph shows that qualitative STS handbooks (green, solid lines) in general show a higher level of similarity with the journal literature. H1-Qual77 and H3-Qual95 exhibit the highest level of similarity but this has been slowly decreasing over time. Quantitative STS exhibits a lower level of similarity between handbooks and journal literature. H2-Quant88, which shows somewhat higher similarity in almost all time periods, was produced with the intent of integrating somewhat disparate research efforts in the field of quantitative studies of science and technology. Most importantly, we see no changes in the level of similarity between handbooks and articles around the times when handbooks were published. This suggests that the role of handbooks in focusing research efforts is very small, at least in STS. The same methodology can be employed to examine the relation of genres in other disciplines.

## 4 Discussion and conclusions

This study has confirmed, using quantitative methods, the differentiation in science and technology studies along methodological lines into qualitative and quantitative STS. We found, based on the analysis of the similarity of words used in the five STS handbooks, that handbooks split into qualitative (H1-Qual77, H3-Qual95, and H5-Qual08) and quantitative (H2-Quant88, H4-Quant05). Furthermore, we show higher similarity in the most recent two qualitative handbooks with respect to the first one. This begs the question of whether the qualitative and quantitative handbooks represent two sides of the same domain or are differentiated enough to be considered as separate research areas. The position of the first handbook in the qualitative tradition is very interesting given the commonly held view that because one of its editors was Derek de Solla Price, one of the founders of scientometrics, the handbook itself would reflect the unity between quantitative and qualitative approaches and thus be an example of a work showing the common origin of the two traditions. Yet this is not reflected in our analysis, which shows that the similarity between this first handbook and the quantitative ones is as low as of the subsequent qualitative handbooks.

Our study has also shown different roles that handbooks played in STS compared to journal articles. STS handbooks are different from the STS journal literature, in that the handbooks have been written largely by invitation. Therefore the editors had a much more prominent role in shaping the content of the handbooks than the journal editors have in shaping the profile of journal articles. Nevertheless, the handbooks do not represent only the views of a handful of editors. They have had large advisory boards and the backing of major institutions and associations.

Our analysis of the trends in similarity between individual handbooks and aggregate sets of journal articles published in a given year has shown that handbooks in this field did not play a special role in focusing research efforts or introducing their decline. It therefore follows that the handbooks have served a programmatic role in the sense that the editors have made a concerted effort to demarcate the field of STS. In this respect the handbooks are written as much for the external as for internal audiences. In addition, all handbooks have an educational function. Therefore, the editors of the handbooks provided updated chapters for the topics presumably considered to be of lasting importance to the field (e.g., laboratory studies, public understanding of science, science indicators). In addition, topics such as 'developed countries' were covered in handbooks, unlike the journal literature, possibly as a concerted effort by editors to be inclusive.

The goals that the editors of the first two qualitative handbooks tried to achieve–to add to the integration and formation of the field (H1-Qual77) and to map and identify the field (H3-Qual95)–may explain the high level of similarity between these two handbooks and the journal literature throughout the period studied. The higher level of similarity of chapter topics with journal articles that these handbooks have may be the result of choosing the topics that represent the core of the field, such as scientific knowledge, scientific community, technology, and science policy and at the same time stake the territory within the wider ecology of related fields. The problems were obviously already present in the literature, so it does not seem that the handbooks played any role in focusing attention, (e.g., we see no spike in similarity following the publication of a given handbook - gray circle in Figure 5). Also, the topics remained covered in the journal literature, which means that the handbook chapters also did not provide definitive summaries of these topics. This conjunction is further supported by the fact that the level of similarity between H5-Qual08 and journal articles is lower. The editors of this handbook presumably felt that STS has finished with its formative stage and grown into a significant field and thus decided to focus on topics that may highlight both the breadth of the field and relatively new lines of research. Yet, as we see, these topics are not exclusively recent.

The first quantitative handbook had a goal of integrating the literature of quantitative STS. This is probably why it has a higher level of similarity with the journal literature than does the second one, with a goal to simply provide the state-of-the-art in the field. It is notable that while in qualitative STS we see a slight decrease in similarity in the focus of the two genres, in quantitative STS we see a slight increase. This may be the result of quantitative STS being both a younger field that is going through the phase of focus, but also being more coherent and less dispersed. Qualitative handbooks also play no special role in either focusing research efforts or marking the decline of topics, since we see no changes in trends around the times when each of the handbooks was published. Note that the analyses we performed average over all chapters in a given handbook. It is possible that certain specific chapters have helped to focus research published in journal articles. This would be an interesting topic for future study.

In addition to looking at the role that handbooks play in topic creation and diffusion, especially in comparison to journal articles, we also provided detailed analyses of both the major topics covered in handbooks and how they have changed over time as well as the closer look at the individual topics covered in handbooks and journal articles.

We identified particular emphases in each handbook using words and bi-grams as the units of analysis. First, each handbook reflects its grounding in certain time periods—for example, the first handbook (published in 1977) emphasized World War II, atomic energy, international affairs, and foreign policy. By the second handbook (published in 1988), the emphasis was on foreign trade, West Germany, and genetic engineering. Second, no phrase occurs as a frequently used bi-gram in all five handbooks. The most frequently used phrase is "science policy", which occurs in four handbooks, although its usage is in decline in both qualitative and quantitative handbooks. This may be the result of science policy differentiating into a field in its own right. Finally, we identified a number of themes present across all of the handbooks. One such theme is technology, the focus on which has been pervasive despite the change in the phrases emphasized. The top phrases related to this theme include: technological development (H1-Qual77), science technology (H1-Qual77, H4-Quant05, H5-Qual08), technology studies (H3-Qual95), technology policy (H3-Qual95), military technology (H3-Qual95), technology assessment (H3-Qual95), technological opportunities (H4-Quant05), technological determinism (H5-Qual08), and information technologies (H5-Qual08). Other such themes are: (a) international studies, with frequently occurring phrases such as developing countries (H1-Qual77, H4-Quant05), foreign policy (H1-Qual77), international affairs (H1-Qual77), foreign trade (H2-Quant88), and international collaboration (H4-Quant05) and (b) sociality that appears regardless of the qualitative, quantitative divide and is evident by the usage of terms such as scientific community, social worlds, and social studies.

One of the interesting findings of this study is the identification of chapters of shared interest across the qualitative and quantitative divide and the nuanced differences when it comes to studying the topics covered in these chapters: technology, gender and policy. Overall, qualitative STS focuses on the larger context around a phenomenon and problematizing it, while quantitative STS focuses on measures and indicators. Thus, in regards to technology, the emphasis on indicators, measurements and specific technologies in the quantitative handbooks suggest a descriptive approach, contrasted against the socio-critical approach of the qualitative handbooks. A similar distinction is made in the treatment of the subject of gender, with the quantitative approach more focused on measurement than on problematizing the phenomenon. This conclusion is supported by the fact that there is no explicit mention of theory in the quantitative discussion of gender in the quantitative handbook.

In answering the question: what have been the major topics covered in handbooks over time? we have shown that by analyzing only the most frequently used topics in each handbook one can reconstruct the phases in the development of STS discussed in previous studies (Jasanoff, 2010; Martin et al., 2012). For example, despite the fact that the last two qualitative handbooks have had chapters on laboratory studies, our analysis of term usage has shown that the focus on laboratory studies has declined over time. This is not surprising, given that the so-called laboratory studies (e.g., Latour & Woolgar, 1986) were the early examples (in the 1970s and 1980s) of an increased interest of STS scholars in knowledge production in local settings. While the analyses of the rising topics shows continuing interest in "science-as-practice" (Pickering, 1992), laboratory studies have lost their dominance in this line of research. Instead, the steady, if not increased, interest of qualitative STS scholars in knowledge production in local settings can be traced by an increase in mentions of Knorr Cetina, an active and influential researcher of microsociological approaches in general and ethnographic approaches in studying science in particular. Similarly, the increased usage of the term 'social worlds' exemplifies the further development of microsociological approaches to studying science. Among the best known concepts introduced by STS researchers working in this tradition is that of 'boundary objects' (Star & Griesemer, 1989). There has also been an increase in the usage of phrase 'knowledge production'. This phrase has been used in relation to research on epistemic cultures the major proponent of which was Knorr Cetina (1999). However, the increase of interest in 'knowledge production' can also been attributed to the 1990s interest in creating knowledge production models. The two best known ones are: mode1/mode 2 model (Gibbons et al., 1994; Nowotny, Scott, & Gibbons, 2001) and triple helix (Etzkowitz & Leydesdorff, 1997; Leydesdorff & Etzkowitz, 1997).

There is also a range of terms that have seen an increase in usage that are tied to the role of science in the modern world, especially focusing on "understanding of the engagement of science and technology with politics and publics" (Hackett, 2008, p. 429). For example, the terms 'social movement', health movement', and 'public participation' are tied to an increased interest in the new ways of interaction between non-experts with scientific knowledge. This interaction has been particularly visible in medical research and interaction with patient organizations (Bucchi & Neresini, 2008). As Jasanoff (2010) has observed "increasingly…the consequences of global imbalances in S&T innovation, and their implications for human rights and social justice, have emerged as focal points of STS scholarship" (p. 195) and this is something that we have found in our analyses.

Finally, the dynamic nature of technology studies in this domain, demonstrated by the bi-gram analysis, is reinforced by the analysis of declining and rising topics—while many particular technology bi-grams are decreasing, technology phrases are persistent and demonstrate a novel approach to studying technology. Rising and declining topics demonstrate a movement away from examining the technology in the light of its contribution to our society, to a more critical stance in which technological development is not taken for granted and different influences on technology on our society and daily lives are being brought into focus. This change in attitudes is exemplified by the increased usage of the term 'technological determinism'. We also witness an increased interest in 'information technology' and 'computer science', probably prompted by the influence that the Internet and the World Wide Web are having both on the science, but also every other aspect of human lives. In addition, the transformative influence of the new information technology on science is shown by an increased use of term 'e-science' as a particular 'type' of science carried out in the changed technological environment.

Our analysis of topic in handbooks and journal articles also demonstrates different emphases between the genres. This was particularly true in the comparison between the quantitative STS handbooks and corresponding journal articles, where only a third of the top topics were shared. More similarities were seen in the citing profiles between qualitative handbooks/journals and quantitative handbooks/journals. This finding may be somewhat surprising given the known plurality in qualitative STS and often competing schools of thought. However, this may have to do more with the way these two fields develop, especially the speed of the development. While qualitative STS has been involved in sometimes fierce battles (Martin et al., 2012), the range of topics covered has not changed radically. In addition, this analysis suggests that handbooks lag behind journal articles.

An additional finding of this analysis is that the qualitative handbooks and journals showed a decreased use of the terms science studies, science policy, scientific community, and social studies. In a study of the cognitive focus of STS based on the analysis of four journals, van den Besselaar (2000) found that the policy researchers started using bibliometric approaches in the 1980s and thus moved closer to the scientometric community. However, as we demonstrated in the analysis of trending topics in quantitative STS, "science policy" has decreased as a topic in quantitative STS handbooks as well. The closer ties to scientometrics may be obvious in the journal literature. However, due to the lack of abstracts in our dataset at the time of first quantitative handbook (1988), we could not perform a trending topics analysis for quantitative STS journals and confirm van den Besselaar's finding. One possible indicator of more permeable boundaries between quantitative STS and science policy studies can be seen in the significant usage of the major policy journal, *Research Policy,* in both the quantitative STS handbooks and journal articles.

In conclusion, this study has presented various methodologies, some of them novel, which are useful in exploring the relationship between genres. The comparison between two particular genres, handbooks and journal articles, in STS with respect to the roles they play in knowledge creation and dissemination, has shown that in the case of science studies, handbooks play no special role in either focusing the research efforts or introducing their decline. Instead, handbooks primarily play a programmatic role in the way that they demarcate a field and can thus be very useful for educational purposes. The study has also shown that the large-scale analysis of full text of the handbooks provides a rather accurate view, as compared to other sources of evidence, of the development of the field and its rather nuanced internal differences. We believe that this study has set a solid foundation for future studies that will study comparatively additional genres and different disciplines. Such studies will not only enhance our understanding of the roles different genres play in the process of knowledge production and dissemination, but will serve as valuable additional source of evidence for the study of the evolution of different fields.

While the analysis of the full text of five handbooks in this article was rather involved (primarily because of the unavailability of electronic texts), performing detailed studies that elucidate roles of different genres in the changing ecology of scholarly communications is of paramount importance to our field. As handbooks and other books are increasingly indexed by Scopus and Web of Science, and the electronic texts become available, their analysis will become easier over time. Incorporation of these other valuable sources to the study of science will be greatly enhanced if we have good understanding of their role in the knowledge production and dissemination as well as the development of fields.

## Acknowledgments

The authors would like to thank the students who scanned, OCR'ed and cleaned each of the handbooks for analysis (Andrew Tsou, Chaoqun Ni, Chenwei Zhang). This works was supported by the National Science Foundation (grant SMA-1208804) as part of the Digging into Data initiative.